\documentclass{article}
\usepackage{LaThuileFPSpro}
\newcommand{\be}{\begin{eqnarray}}
\newcommand{\ee}{\end{eqnarray}}
\newcommand{\bi}{\bibitem}

\newcommand{\dt}{\partial_t}
\newcommand{\pr}{\partial}
\newcommand{\rotv}{{\bf \nabla}\times {\bf v}}

\begin{document}
\title{ 
MAGNETIC FIELDS IN COSMOLOGY 
  }
\author{
A.D. Dolgov  
\\
  {\em INFN, sezione di Ferrara, Via Paradiso, 12 - 44100 Ferrara, Italy }\\
  {\em and}\\
 {\em ITEP, Bol. Cheremushkinskaya 25, Moscow 113259, Russia} \\
  }
\maketitle

\baselineskip=11.6pt

\begin{abstract}
Different mechanisms which may possibly explain existence of magnetic fields on astronomically large scales are described. A recently suggested model of magnetic fields generation slightly before hydrogen recombination is discussed in more detail.  
\end{abstract}
\newpage
\section{Introduction \label{s-intr}}
It is known from observations, mostly from Faraday effect, that there are magnetic fields in
galaxies with magnitude of a few micro-Gauss and coherence length of the order of
galactic size,   $l_{gal} \sim$ ( a few) kpc, The energy density of such fields is close to 
the energy density of the cosmic microwave background radiation (CMBR):
\be 
\rho_B = B^2/8\pi \sim \rho_{\gamma} \sim 10^{-4} \rho_m 
\approx 3\cdot 10^{-34}\,{\rm g/cm}^3
\label{rho-B}
\ee
where $\rho_{\gamma}$ is the energy density of CMBR and
$\rho_m \approx 1.5$ keV/cm$^3$ is the total cosmological mass/energy density.
Though the magnitude of such fields is small in comparison with the fields of stars or
even planets (see below) it is very difficult to understand their huge coherence length.
Even more puzzling is a possible existence of intergalactic magnetic fields which are
3 orders of magnitude weaker but coherent at a Megaparsec scale, for a review see 
ref.\cite{magn-rev}. 

For comparison, magnetic field of the earth is $B_\otimes = 0.5$ G, magnetic fields of
solar type stars can be as large as $10^3$ G, magnetic fields of white dwarfs may reach
$10^9$ G, and absolute champions are neutron stars with magnetic fields at the level
of $10^{13}$ G.

Possible creation mechanisms of large scale galactic or intergalactic fields can be roughly 
separated into four classes (for reviews see ref.\cite{gnrtn}):
\begin{enumerate}
\item{}
Conventional (astrophysical) mechanism based on stellar ejecta of magnetic fields with subsequent magnetic lines reconnection (from different stars) to create a homogenous on large scale field component. This mechanism was reviewed recently in ref.\cite{biermann}
\item{}
Processes in the early universe which invoke inflation to stretch the characteristic scale of
the field up to galactic or even larger scales. A  comprehensive list of references on different versions of these mechanisms can be found in the recent paper\cite{dimopoulos02}.
\item{}
Phase transitions in the early universe during which strong magnetic fields could be created 
but at a very small scale. 
\item{}
Relaxation of previously created inhomogeneities. In this process turbulent or laminar flow
of primeval plasma with non-zero vorticity  could be generated and due to different mobilities
of charge carriers vortical electric currents producing magnetic fields would be created. Such
mechanism may operate either in the early or relatively late universe.
\end{enumerate}
All these mechanisms either generate magnetic fields at very small scales or of  insufficiently 
large amplitude. In the first case "Brownian" type reconnection of magnetic field lines is
necessary for creation of  coherent magnetic fields at large distances. The amplitude of the field
in the course of reconnection decreases as 
$(l_{in}/l_{fin})^{3/2}$ where $l_{in} $ and $l_{fin}$ are the
initial and final coherence lengths. Inflationary stretched fields are usually rather weak,
though may have a very large coherence length. In both cases galactic dynamo\cite{dyn}
should amplify the seed magnetic fields up to the necessary magnitude.

\section{A comment on astrophysical mechanism \label{s-astro}}

The total energy of galactic magnetic field in a large galaxy, e.g. in the Milky Way, is equal to:
\be 
{\cal E}_{gal}^B = \frac{4}{3} \,\pi R_{gal}^3 \rho_B \approx  10 M_\odot
\label{E-gal}
\ee
where $M_\odot$ is the solar mass. Magnetic energy of a neutron star with radius 
$R_{ns}=10^6$cm
is equal to ${\cal E}_{ns}^B \approx 10^{-11} M_\odot (B/10^{13}G)^2$. Magnetic energy 
of white dwarfs with $R_{wd} = 10^9$cm is 
${\cal E}_{wd}^B \approx 10^{-10} M_\odot (B/10^{9}G)^2$. Thus about $10^{11}$ white dwarfs
with magnetic field $B=10^{10}$G or $10^{12}$ neutron stars with $B=10^{13}$G 
should be in the Galaxy to feed galactic magnetic field if the
energy  of the field were not lost in the process of line reconnection. Since the total number of
stars in the Galaxy is about $2\cdot 10^{11}$ it is hardly possible to meet this requirement. 
Of course this estimate is quite rough and more accurate considerations may be not so
negative but still it shows that efficiency of stellar creation of galactic magnetic field should
be very high to satisfy energy  constraints.

\section{Generation of seed magnetic fields in the early universe}

It is well known that during inflation very long gravitational waves and large scale scalar field
perturbations are generated. A natural question is: "why not electromagnetic fields?" 
The answer is that scalars and tensor fields are not conformally invariant even in the zero
mass case, while photons are. Conformal invariance means that rescaling metric
with an arbitrary factor $b (t,{\bf r})$ and the fields by the same factor to a power
depending on the spin of the field
we arrive to formally the same action written in terms of new variables.
According to the Parker theorem\cite{parker68} conformally invariant fields
are not generated in conformally  flat space-time. Indeed it is known that cosmological 
Friedman-Robertson-Walker  (FRW) background
is conformally flat, i.e. after the proper coordinate choice the metric can be written in the form
\be
ds^2 = a(\tau, {\bf x})^2\left( d\tau^2 - d{\bf x}^2\right),
\label{conf-flat}
\ee
Thus after conformal transformation with the factor $b=a(\tau, {\bf x})$ to a proper power we can exclude FRW-gravity for conformally invariant fields. Fortunately, as it has been already mentioned, this cannot be done for scalar (in particular, inflaton) fields. Otherwise cosmological density perturbations would not be generated and we would not be here. 

Several possible ways to break conformal invariance of electrodynamics and to create seed
magnetic fields have been discussed in the literature:
\begin{enumerate}
\item{}
New non-minimal interaction of electromagnetic field with gravity, possibly not gauge 
invariant\cite{turner88} :
\be
{\cal L} = C_1 R A_\mu A^\mu + C_2 R_{\mu\nu} A^\mu A^\nu + 
C_3 R_{\mu\nu\alpha\beta} F^{\mu\nu}F^{\alpha\beta} +...
\label{ltw}
\ee
where $A_\mu$ is the electromagnetic vector-potential, $F^{\mu\nu}$ is the Maxwell tensor,
$R_{\mu\nu\alpha\beta} $ is the Riemann tensor, $R_{\mu\nu}$ is the Ricci tensor, and $R$ is
the curvature scalar. 
\item{}
Interaction with a new hypothetical field, dilaton, $\theta$\cite{ratra}:
\be
{\cal L} = -(1/4)\,e^\theta\, F_{\mu\nu}F^{\mu\nu}
\label{lr}
\ee
\item{}
Quantum conformal anomaly due to famous triangle diagram which leads to non-zero
trace of the energy-momentum tensor of electromagnetic field and breaks in this way 
conformal invariance of electrodynamics\cite{dolgov93}:
\be
T_\mu^\mu = \kappa  F_{\mu\nu}F^{\mu\nu}
\label{tmumu} 
\ee
where the constant coefficient $\kappa$ depends upon the rank of the gauge group and the number of charged fermions. For $SU(N)$ with $N_f$ number of charged fermions it is:
$\kappa = (\alpha/\pi) (11N/3-2N_f/3)$.
\end{enumerate}

In all these models electromagnetic waves could be generated at inflationary stage and 
sufficiently large magnetic fields at very large scales would be created to serve as seeds 
for galactic fields, if one takes  appropriate values of of coupling constants in the Lagrangians
or sufficiently large number of charge particles in the third case.
Unsatisfactory features of this approach are introduction of new fields
or interactions, though the second case looks quite natural in string inspired theories, while
the third one looks good in grand unification models especially keeping in mind that the fine
structure coupling constant $\alpha$ becomes quite large, about 1/40, at the unification 
scale. The latter makes the mechanism much more efficient.

\section{Generation of vorticity perturbations by cosmological inhomogeneities \label{s-vort} }

There are several mechanisms which might create inhomogeneities in the primeval
plasma whose relaxation could lead to generation of primordial magnetic fields:
\begin{enumerate}
\item{}
First order phase transitions creating bubbles of one phase inside another\cite{pt}.  Though
the magnitude of magnetic fields produced on the boundaries between the phases 
could be very large, the characteristic scale is extremely small and it is difficult to stretch 
it up to galactic size.  
\item{}
Creation of stochastic inhomogeneities in cosmological charge 
asymmetry, either electric\cite{dolgov-s93}, or e.g. leptonic\cite{dolgov-gr01} at 
large scales which produce turbulent electric currents and, in turn, 
magnetic fields. The first of these models could be quite efficient but it demands rather
unusual physics. For realization of the second model only one rather innocent assumption is necessary, namely, an existence of sterile neutrino, $\nu_s$, very weakly mixed with active 
ones, $\nu_{e,\mu,\tau}$. If $\nu_s$ is lighter than $\nu_a$ and the mass difference is
sufficiently large then the MSW-resonance transition between $\nu_s$ and active
neutrinos would take place giving rise to a large, about 0.1, and strongly fluctuating lepton asymmetry in the active neutrino sector\cite{di-bari}. When the wave length of the domain
with a large lepton (or anti-lepton) number crossed horizon, the neutrino or antineutrino
flux from this domain would induce electric current by scattering on electrons or positrons
because of different $\nu e^-$-  and $\nu e^+$-cross-sections. In this process hydrodynamical
flows with large Reynolds numbers would be generated and turbulent vortical currents could be 
produced. The characteristic wave length of the generated magnetic field is about 100 pc
and chaotic line reconnection and large but not unreasonable dynamo amplification are
necessary for explanation of the observed galactic fields.
\item{}
Generation of seed magnetic fields slightly before hydrogen recombination epoch\cite{bedo}
through relaxation of the usual density perturbations which are known from observations
to exist. This model does not demand any 
new physics and predicts quite promising amplitude of seed fields at galactic scales. 
Since the work\cite{bedo} is new, not yet published, we will discuss it
in the next section in some detail.
\end{enumerate} 

\section{Generation of large scale magnetic fields at recombination epoch}

We will consider the period when the universe was already quite old and cool with 
temperature about 1-100 eV, i.e. somewhat before hydrogen recombination.  The usual cosmological density perturbations are known to exist at that time, with rather small
amplitude, $\delta\rho/\rho \sim 10^{-4}$.  The motion of the cosmological plasma, or better 
to say fluid, under pressure forces is governed by the hydrodynamical
equation (see e.g. the book\cite{ll-hydro}):
\be
\rho \left(\dt v_i + v_k \pr_k v_i \right) =
-\pr_i p + \pr_k \left[\eta\,\left(\pr_k v_i + \pr_i v_k - 
{2\over 3} \delta_{ik} \pr_j v_j \right) + \pr_i \left( \zeta\, \pr_j v_j
\right) \right] 
\label{master-eq}
\ee
where $v$ is the velocity of the fluid element, $\rho$ and $p$
are respectively the energy and pressure densities of the fluid, 
and $\eta$ and $\zeta$ are the first and second viscosity coefficients.
In the case of constant viscosity coefficients this equation is reduced to the
well known Navier-Stokes equation. The coefficient $\eta$ is related to the
mean free path of particles in fluid as 
\be
\eta /\rho \equiv \nu = l_{f}
\label{lmfp}
\ee 
In what follows we disregard the second viscosity $\zeta$.

The behavior of the solution to eq.~(\ref{master-eq}) crucially 
depends upon the value of the Reynolds number
\be
R_\lambda = \frac{v \lambda}{\nu} 
\label{R}
\ee
where $\lambda$ is the wavelength of the velocity perturbations.
If $R\gg 1$, then the fluid motion would become turbulent and non-zero 
vorticity would be created by spontaneously generated turbulent eddies. 
In the opposite case of low $R$ the motion is smooth and 
and no vorticity could be generated in the first order in $\delta \rho$.
One would expect that in the case of scalar perturbations vorticity remains zero
in any order in $\delta\rho$. However, this is not the case, as we will argue in what
follows.

Let us first estimate the Reynolds number of the fluid motion created 
by the pressure gradient in eq. (\ref{master-eq}). 
To this end we assume that the liquid is quasi incompressible and
homogeneous, so that the second term in the r.h.s. of this equation 
can be neglected. This is approximately correct and the obtained 
magnitude of the fluid velocity is sufficiently accurate. 
In this approximation eq. (\ref{master-eq}) reduces to a much 
simpler one:
\be
\dt {\bf v} + \left( {\bf v}\, {\bf \nabla} \right) {\bf v} -
\nu \Delta {\bf v} = -{{\bf \nabla} p \over \rho } 
\label{ns-eq}
\ee

A comment worth making at this stage. The complete system of equations
includes also continuity equation which connects the time variation of 
energy density with the hydrodynamical flux (see below eq.~(\ref{contin}))
and the Poisson equation for gravitational potential induced by density 
inhomogeneities. We will however neglect the gravitational force and the
back reaction of the fluid motion on the density perturbation. This
approximation would give a reasonable estimate of the fluid velocity
for the time intervals when acoustic oscillations are not yet developed,
i.e. for $t<\lambda/v_s$, where $\lambda$ is the wave length of the 
perturbation and $v_s$ is the speed of sound (in the case under 
consideration $v_s^2 = 1/3$). In fact the wave length should be larger
than the photon mean free path, to avoid diffusion damping, and 
the characteristic time interval, as we see below, should be  
somewhat larger than $\lambda$. So we may hope that our estimates of the
velocity are reasonable enough. Neglecting gravitational forces, especially
those induced by dark matter would result in a smaller magnitude of the
fluid velocity, so the real effect should be somewhat larger.

For small velocities (or sufficiently small wavelengths) we may neglect
the second term in the l.h.s. with respect to the third one. 
In this approximation the equation becomes linear and can be easily solved 
for the Fourier transformed quantities. 
Assuming that the parameters are time-independent (though it
is not necessary) we obtain: 
\be
{\bf v_k} =- {i{\bf k} \over 3k^2 \nu}\,\delta_k\,
\left[ 1 - \exp (- \nu k^2 t) \right] 
\label{v0}
\ee
where $\delta_k = (\delta \rho/\rho)_k$ is the Fourier transform of 
relative density perturbations, $\delta \rho /\rho$; 
its natural value is $\sim 10^{-4}$, though it might be much larger 
at small scales. The coefficient $1/3$ comes from
equation of state of relativistic gas, $p=\rho/3$.

Therefore, for the Reynolds number we obtain:
\be
R_k = \frac{\delta_k}{3 \left( k\,\nu \right)^2 } 
\left[1 -\exp \left( - {\nu k^2 t } \right) \right]  \, . 
\label{R1}
\ee
If $\delta_k$ is weakly dependent on $k$, then $R_k$
is a monotonically rising function of the wavelength $\lambda = 2\pi/k$.  
For $t \ll \lambda^2/\nu $ it takes the value  
\be
R_k  = { t \over 3\nu} \, \delta_k \ll 1,
\label{R-max}
\ee 
so the hydrodynamical flow remains laminar and vorticity is not spontaneously generated.
Dynamical generation of vorticity is governed by the equation:
\be
\dt {\bf \Omega} - \nu \Delta\, {\bf \Omega} = 
- {\bf \nabla} \times \left({ {\bf\nabla} p \over \rho }\right)
\label{dt-omega}
\ee
where ${\bf \Omega} = \rotv$ and we assume that velocity is small 
so that the term quadratic in $v$ was neglected. 
If the r.h.s. is non-vanishing, then
${\bf \Omega}$ would be non-zero too. However usually pressure density is 
proportional to the energy density, $p = w\rho$, with a constant
coefficient $w$ and hence
${\bf \nabla} \times \left({ {\bf \nabla} p / \rho }\right) =0$.We can see that this is not so because
cosmic plasma consist of different components whose motion is somewhat different. Let us 
assume that plasma is in local thermal equilibrium with common temperature $T({\bf x})$.
This assumption is justified by a large interaction rate between radiation and charged particles. 
If $T$ would be the only parameter
which determines the state of the medium, then vorticity would 
not be generated
because we would have in our disposal only ${\bf\nabla} T$ and 
it is impossible to construct non-vanishing $\rotv$ from the
gradient of only one scalar function.
However, distributions of charged particles depend upon one more function,
their chemical potential: 
\be
f = \exp \left[ -{E\over T({\bf x})} + \xi ({\bf x}) \right]
\label{f}
\ee
where the dimensionless chemical potential 
$\xi$ can be readily expressed through
particle number density $n_e \approx n_B = \beta (x) n_\gamma$ with 
$\beta (x) = 6\cdot 10^{-10} + \delta \beta (x)$:
\be
\xi (x) = \ln \beta (x) + {\rm const} 
\label{xi-x}
\ee
Hence we will find that the source term in the vorticity 
equation (\ref{dt-omega}) is equal to:
\be
S_k\equiv -\epsilon_{ijk}\, \pr_j \left({\pr_i p \over \rho}\right) =
\epsilon_{ijk}\, { \pr_i \rho_\gamma \over 3 \rho_{\rm tot}}\,
{\pr_j \beta \over \beta}\, {\rho_b \over \rho_{\rm tot}} 
\label{source}
\ee
An essential feature here is that the spatial distribution of charged 
particles does not repeat the distribution of photons and hence the
vectors ${\bf \nabla} \rho_\gamma$ and  ${\bf \nabla} \beta$ are not 
collinear.  This could occur if, for the wave length
corresponding to subgalactic scales,
there exist baryon isocurvature fluctuations and thus
$\rho(x)$ and $\beta(x)$ have different profiles.
As we have mentioned above, different mean free paths of
photons and charged particles would maintain such non-collinearity of
the order of unity at the scales $\lambda \sim l_{\gamma} $.  
Moreover, even in the case of adiabatic perturbations
a shift in the distribution of photons and charged particles 
could also be created 
because of acoustic oscillations that proceeded with different phases of
radiation and matter densities. At the scales 
$\lambda \leq l_{\gamma}$
perturbations in the the plasma temperature would be erased by the
diffusion damping\cite{silk68}, while for $\lambda \gg l_{\gamma}$
the diffusion processes are not efficient and one would expect self-similar
perturbation leading to collinearity of 
${\bf \nabla} \rho_\gamma$ and  ${\bf \nabla} \beta$. On the other hand,
when $\lambda$ entered under horizon acoustic oscillations begun which
destroyed the self-similarity.
Thus the expected wavelengths of vorticity perturbations should be
between $l_{\gamma} <\lambda < H^{-1}$.

Surprisingly vorticity can be also generated (and in the case under consideration even
a larger one) if   perturbations in plasma are
determined by a single scalar function, for example, by $T(t,{\bf x})$
because it might be proportional to the product 
$\partial_i T(t,{\bf x})\,\partial_j T(t',{\bf x}) $.  These two gradients
generally are not collinear if taken at different time moments $t$ and
$t'$. To see that, let us start from the Boltzmann equation for
the distribution function $f(t,{\bf x},E,{\bf p})$ of photons:
\begin{equation}
\left( \frac{\partial}{\partial t} + {\bf V}\cdot{\bf \nabla} -
H\, {\bf p} \frac{\partial}{\partial {\bf p}} +
{\bf F}\, \frac{\partial}{\partial {\bf p}  } \right) 
f(t,{\bf x},E,{\bf p})
= I_{\rm coll}\left[f_{a}, f_b,...\right] ~, 
\label{boltz}
\end{equation}
where ${\bf V} = {\bf p}/E$ is the particle velocity (not to be confused
with the velocity ${\bf v}$ of macroscopic motion of the medium,
for photons $V=1$, while $v \ll 1$),
$E$ and ${\bf p}$ are respectively the particle energy and
spatial momentum, $H$ is the universe expansion rate, ${\bf F}$ is
an external force acting on particles in question (the latter is assumed
to be absent), and $I\left[f_{a}, f_b,...\right]$ is the collision 
integral depending on the distributions $f_a$ of all participating 
particles.

At temperatures in eV-range only the Thomson scattering of photons on 
electrons is essential, so the collision integral is dominated by the
elastic term. Integrating both parts of eq.~(\ref{boltz}) over 
$d^3 p/(2\pi)^3$ we arrive to the continuity equation:
\be
\dot n ({\bf x}) + {\bf \nabla}{\bf J} = 0
\label{contin}
\ee
where ${\bf J}$ is the photon flux given by
\be
{\bf J } \equiv {\bf v} n = \int {d^3 p \over (2\pi)^3}\,
{{\bf p}\over E}\, f
\label{J}
\ee
and ${\bf v}$ is the average macroscopic velocity of the photon plasma.
Using the standard arguments one can derive from eq.~(\ref{contin}) the
diffusion equation:
\be
\dot n = D\, \Delta n
\label{difus}
\ee
where $D\approx l_{\gamma}/3$ is the diffusion coefficient.
We will use this equation below to determine time evolution of the photon
temperature $T$.

If the elastic reaction rate
$\Gamma_{\rm el} = \sigma_{\rm Th} n_e X_e = 1/l_{\gamma}$ 
is sufficiently large, local thermal equilibrium would be established
and the photon distribution would be approximately given by
\be
f \approx f_{0} = 
\exp \left(- E/T +\xi \right)
\label{f0}
\ee
where the temperature and effective chemical potential could be
functions of time and space coordinates: $T=T(t,{\bf x})$ and
$\xi=\xi(t,{\bf x})$, and the photon mean free path is given by
$ l_\gamma = 30\,{\rm pc}\,/X_e(T) T^3_{eV}$, where $T_{eV}$ is the plasma 
temperature in eV and 
 $X_e$ is a fraction of the free electrons: $X_e(z)$ is practically 
1 for $z > 1500$, and sharply decreases for smaller $z$'s, reaching 
values $\sim 10^{-5}$ at $z<1000$.

Evidently $f_0$ annihilates the collision 
integral. We can find correction to this distribution, $f =f_0 + f_1$,
substituting this expression into kinetic equation (\ref{boltz}) and 
approximating the collision integral in the usual way as 
$-\Gamma_{\rm el} f_1$:
\be
\left( K  + \Gamma_{\rm el}\right) f_1 = - K f_0
\label{Df1}
\ee
where $K$ is the differential operator, 
$K = \partial_t + \left({\bf V} \,{\bf \nabla}\right)$.
The solution of this equation is straightforward:
\be
f_1 \left(t,{\bf x},E,{\bf V}\right)=                
- \int_0^t d\tau_1 
\exp{ \left[ -\int_0^{\tau_1} d\tau_2 
\Gamma_{el} \left(t-\tau_2, {\bf x}-{\bf V}\tau_2 \right)\right]}  \nonumber \\
K f_0\left(t-\tau_1, {\bf x} -{\bf V}\tau_1
\right)
\label{f1}
\ee
Using this result we can calculate the average macroscopic
velocity of the plasma. The calculations are especially simple if
elastic scattering rate is high and the integrals are dominated by
small values of $\tau_1$. In this case we obtain: 
\be
{ v_j}(t,{\bf x}) = {\int d^3 p { V_j} f_1(t,{\bf x},E,{\bf V})
\over \int d^3 p f_0(t,{\bf x},E) }
\label{v}
\ee
and the vorticity, $\Omega_i=\epsilon_{ijk} \partial_j v_k$ is:
\be
\Omega_i \approx 6\epsilon_{ijl}\, l_{\gamma}^2\, 
\left({\partial_j T \over T}\right)\, 
\left({\partial_l \partial_t T \over T}\right)
\label{omega-i}
\ee

To estimate time derivatives of the temperature we will use the diffusion 
equation~(\ref{difus}), from which we find $\partial_t T = D \Delta T$
and finally obtain for vorticity with the wave vector $k=2\pi/\lambda$:
\be
|{\bf \Omega}|_\lambda 
\approx 2\, \left({\delta T \over T}\right)^2_\lambda\,
l_{\gamma}^3\,k^4 \approx 3\cdot 10^3
\left({\delta T \over T}\right)^2_\lambda\
{l_{\gamma}^3\over \lambda^4}
\label{omega-fin}
\ee
Since the photon diffusion erases temperature fluctuations at the scales
$\lambda < l_\gamma$ the vorticity reaches maximum value near 
$\lambda \sim l_\gamma$. This magnitude of vorticity is considerably
larger than found previously  with the source term (\ref{source}) and we will rely on it in
the estimates of magnetic field presented below.

Since the conductivity of cosmic plasma is very high, 
\be
\kappa =(3/2 \alpha)\,(n_e / n_\gamma)\,(m_e^{2}  T),
\label{kappa}
\ee
the generation of magnetic field by the
source currents, created by the cosmological inhomogeneities, 
is governed by the well know equation of magnetic hydrodynamics:
\be
\dt {\bf B} = {\bf \nabla} \times \left( {\bf v} \times {\bf B} \right) + 
{1\over \kappa}\, {\bf \nabla} \times {\bf J}
\label{dt-B}
\ee
The electric current $J$ induced by the
relaxation of the density inhomogeneities
would contain two components: electronic and protonic. However
the first one is surely dominant because it is much easier to drift 
electrons than heavier protons. This is why a non-zero
current can be induced in electrically neutral medium. Of course the motion of 
electrons would not produce any excess of electric charge because 
the current could be realized by the flow of the dominant homogeneous part of electron
distribution.

The solution of eq. (\ref{dt-B}) can be roughly written as
\be
B\sim \int_0^t dt_1 \left({2\pi\,J \over \lambda\,\kappa}\right)\,
e^{ 2\pi\,v\,t_1 /\lambda }
\label{b1}
\ee  
An estimate of magnetic field without pregalactic dynamo enhancement
can be easily done if the helical source current is known,
${\bf \nabla \times J} = e n_e {\bf \Omega}$. With 
${\bf \Omega}$ given by
eq.~(\ref{omega-fin}) and $B$ by eq.~(\ref{b1}) we obtain:
\be
{B_0\over T^2}= 0.24\cdot 10^3\left( 4\pi\alpha\right)^{3/2}\,
\left({t\over \lambda}\right)\,
\left({l_\gamma \over \lambda}\right)^3\, 
\left({ T \over m_e }\right)^2 \approx 
10^{-8} T_{\rm eV}^3
\label{B/T2}
\ee
where we took the wavelength equal to the photon mean free path,
$\lambda = l_\gamma$.

If we take into account that linear compression of pregalactic medium 
in the process of galaxy formation is approximately $r\sim 10^2$, 
the seed field in a galaxy after its formation would be 
$r^2 B_0$, i.e. 4 orders 
of magnitude larger than that given by eq.~(\ref{B/T2}) and, for
$T= 1$ eV, a relatively
mild galactic dynamo, about $10^{4}$, is necessary to obtain 
the observed galactic 
magnetic field of a few micro-Gauss at the scale 
$l_B \sim (100/r)$ kpc $ =1$ kpc. 
The seed magnetic fields formed earlier (at 
higher $T$) would have larger magnitude ($\sim T^3$) but their 
characteristic scale would be smaller by factor $1/T^2$. Chaotic line 
reconnection could create magnetic field at larger, galactic scale 
$l_{gal}$, but the 
magnitude of this field would be suppressed by Brownian motion law -
it would drop by the factor $(l_B/l_{gal})^{3/2}$. It is interesting
that according to these results all scales give comparable contributions
at $l_{gal}$. This effect may lead to an enhancement of the field but 
it is difficult to evaluate the latter.
Let us also note that magnetic fields generated by the discussed
mechanism at the cluster scale, 10 Mpc, should be not larger than
$10^{-8}$ $\mu$G if no additional amplification took place.

Larger density perturbations could be helpful for generation of larger 
magnetic field for which dynamo might be unnecessary. Though much bigger
$\delta T$ is not formally excluded at the scale about 100 kpc, but 
to have them at the level $(\delta T/T)^2\sim 10^{-4}$ seems to be too
much. A natural idea is to turn to a later stage, to onset of structure 
formation when $\delta \rho/\rho$ becomes larger than $10^{-2}$. With such
density perturbations strong enough magnetic fields might be generated 
without dynamo amplification. However after recombination the number 
density of charge carriers drops roughly by 5 orders of magnitude.
Correspondingly $l_\gamma$ rises by the same amount and the strength of
the seed field would be 5 orders of magnitude smaller if density 
perturbations and the temperature of formation remained the same. However
both became very much different. Density perturbations rose as
scale-factor,
$(\delta\rho /\rho)^2\sim (T_{eq}/T)^2$, where $T_{eq}\sim 1$ eV is the 
temperature when radiation domination changed into matter domination and 
density perturbations started to rise. Since, $B/T^2\sim T^3$, according
to eq.~(\ref{B/T2}), the net effect of going to smaller $T$ is a decrease
of $B/T^2$ which would be difficult to cure even by later reionization. 
Still, as argued in ref.\cite{langer03}, magnetic field generation,
driven by anisotropic and inhomogeneous radiation pressure (and in this
sense similar to our mechanism) at the epoch of reionization, could end 
up with the field of about $8\cdot 10^{-6} \mu{\rm G}$. This result is 
8 orders of magnitude larger than that found in the earlier 
papers\cite{reion} and quite close to ours (\ref{B/T2}), though these 
two mechanisms operated during very different cosmological epochs and 
were based on different physical phenomena. 

Generation of magnetic field at recombination was also considered in ref.\cite{hogan}
where a much weaker result was found. This difference can be possibly attributed to the 
following effects. We considered above an earlier period when the
photon mean free path was much smaller than the horizon. It gives a factor
about $10^3$ in fluid velocity, eq.(\ref{v0}). Moreover, since in our
case the electrons are tightly bound to photons the electron-photon
fluid moves as a whole (while protons and ions are at rest) and the
electric current induced by macroscopic motion/oscillations of plasma
is noticeably larger.

\section{Conclusion}

Despite many suggested models, the origin of galactic magnetic fields and, especially,
intergalactic, if existence of the latter is confirmed, remains mysterious.   One class of 
models is based on known physics and do not invoke any ad hoc assumptions for the
explanation of the phenomenon. The explanation based  on stellar ejecta 
possibly encounters serious difficulties because of energy constraints. 
The mechanism of field generation just before hydrogen recombination looks reasonably 
good but probably it cannot explain both
galactic and intergalactic fields within the frameworks of the standard cosmology with flat
spectrum of density perturbations. 

Another class of models is based on physical phenomena in the early universe and
its different members are spread between inflationary stage to relatively late  MeV-epoch.
These models manipulate with unknown physics (except possibly the MeV one) and because
of that are much less restricted in their possibilities. It is a difficult task to understand what
mechanism is indeed responsible for creation of the observed magnetic fields. A critical
test would be a possibility of simultaneous explanation of galactic and intergalactic fields. 
To this end a confirmation of a possible existence of the latter is of primary importance.

\section{Acknowledgements}
The hospitality of the Research Center for the Early Universe of the University of Tokyo, where 
this contribution was prepared for publication, is gratefully acknowledged.

\end{document}